\documentclass[fleqn,usenatbib,usedcolumn]{mnras}
\usepackage{mathptmx}

\usepackage[T1]{fontenc}
\usepackage{ae,aecompl}

\usepackage{graphicx}	
\usepackage{amsmath}	

\newcommand{\um}{$\mu$m}

\title[MIR Polarimetry: a Bayesian approach]{Dichroic polarization at mid-infrared wavelengths: a Bayesian approach}

\author[Lopez-Rodriguez et al.]{
E. Lopez-Rodriguez$^{1,2}$\thanks{E-mail: \href{mailto:enrique.lopezrodriguez@utsa.edu}{enrique.lopezrodriguez@utsa.edu}}
\\
	$^{1}$Department of Physics \& Astronomy, University of Texas at San Antonio, One UTSA Circle, San Antonio, TX 78249, USA \\
	$^{2}$Department of Astronomy, University of Texas at Austin, 1 University Station C1400, Austin, TX 78712, USA 
	}

\date{Accepted XXX. Received YYY; in original form ZZZ}

\pubyear{2015}

\begin{document}
\label{firstpage}
\pagerange{\pageref{firstpage}--\pageref{lastpage}}
\maketitle

\begin{abstract}

A fast and general Bayesian inference framework to infer the physical properties of dichroic polarization using mid-infrared imaging- and spectro-polarimetric observations is presented. The Bayesian approach is based on a hierarchical regression and No-U-Turn Sampler method. This approach simultaneously infers the normalized Stokes parameters to find the full family of solutions that best describe the observations. In comparison with previous methods, the developed Bayesian approach allows the user to introduce a customized absorptive polarization component based on the dust composition, and the appropriate extinction curve of the object. This approach allows the user to obtain more precise estimations of the magnetic field strength and geometry for tomographic studies, and information about the dominant polarization components of the object. Based on this model, imaging-polarimetric observations using two or three filters located in the central 9.5$-$10.5 \um, and the edges 8$-$9 \um~and/or 11$-$13 \um, of the wavelength range are recommended to optimally disentangle the polarization mechanisms. 
\end{abstract}





\section{Introduction}
\label{INTRO}

Polarization arising by asymmetric dust grains aligned with the magnetic field in the interstellar medium (ISM) has been probed from the diffuse medium in the ultraviolet, heavily obscured sightless in the near-infrared (NIR), and dense clouds heavily embedded sources using far-IR (FIR) \citep[e.g.][]{Anderson:1996aa,Dotson:2000aa,Clemens:2012aa}. These observations show that the dust grains in the ISM are magnetized and provide an observational method to study the geometry and strength of the magnetic fields \citep[e.g.][]{Davis:1951aa,Chandrasekhar:1953aa,Lazarian:2007aa,Hoang:2015aa}. In general, the short axes of dust grains become aligned along the constant component of the ambient magnetic field producing a differential extinction and thus, a measurable polarization; a polarization mechanism called dichroism. In general, the polarization angle by dichroic absorption is in the direction of the magnetic field onto the plane of the sky, while the polarization angle by dichroic emission is orthogonal to it. The geometry and strength of magnetic fields can be studied through the extinction and/or emission of aligned dust grains.

Mid-IR (MIR; 8$-$13 \um) spectro-polarimetric observations of 55 objects (e.g. young stellar objects, star formation regions, and active galactic nuclei) using 4-m class telescopes found that 90 per cent of the objects can be explained by dichroism \citep{Smith:2000aa}. Although other mechanisms of polarization may be present, these are less dominant at MIR wavelengths. For instance, the polarization efficiency by dust scattering has a steep decrease with wavelength ($\propto \lambda^{-4}$ or $\lambda^{-1}$), while electron scattering is wavelength-independent. At MIR, both mechanisms are mostly extinguished and/or their steep decrease with wavelength makes them negligible. Synchrotron radiation has a constant polarization angle perpendicular to the direction of the magnetic field with a high degree, $>$10\%, of polarization \citep[e.g.][]{Lopez-Rodriguez:2014aa}. Dichroism has a degree of polarization $<$10\% \citep{Smith:2000aa} and the polarization angle can show some wavelength dependence. Dichroic extinction and emission can compete at a single wavelength. Both mechanisms need to be disentangled through a multi-wavelength study. Fortunately, some dust grain features, such as silicates, are present in the MIR wavelength range and both mechanisms can be distinguished. In the case of silicates, the absorptive polarization shows a peak at approximately 10 \um, while the emissive polarization is less structured. In general, if a rotation of the polarization angle with wavelength is observed, then more than one mechanism of polarization may be present. Unless the several polarization mechanisms have intrinsically the same polarization angle, the net polarization angle will be a function of wavelength. Both the degree and angle of polarization profiles are crucial to investigate the emissive and absorptive polarization components at MIR wavelengths.

MIR polarimeters in 8-m class telescopes have shown the potential to investigate magnetic fields in star formation regions \citep{Barnes:2015aa} and active galactic nuclei \citep[][2015 submitted]{Packham:2007aa,Lopez-Rodriguez:2014aa}. For instance, \citet{Barnes:2015aa} and Lopez-Rodriguez et al. (2015, submitted) show a decomposition of the dichroic extinction and emission onto the plane of the sky for K3-50 and NGC 1068, respectively. These authors followed the analysis techniques put forward by \citet{Aitken:2004aa}. This analysis uses the 10 \um~spectral signature of silicates to disentangle the emissive and absorptive polarization components. This method is optimized to minimize $\chi^{2}$ of the fit to the normalized Stokes parameters $q$ and $u$, independently. This technique provides a unique best-fit of the observations regardless of whether or not a unique solution exists, nor whether the Stokes parameters are dependent on each other. Although this approach has been successfully applied to a variety of objects \citep{Smith:2000aa}, the method does not address the use of different extinction curves, which may change for every object, nor does it address other absorption polarization components based on different dust composition. Thus, a fitting procedure giving the full family of possible solutions to explain a set of observations, which 1) simultaneously fits the normalized Stokes parameters, and 2) allows the user the possibility of using different extinction curves and different absorption components based on every object, are needed. This paper aims to present a fast and general Bayesian inference framework to infer the physical properties of the dichroic polarization using MIR imaging- and spectro-polarimetric observations.



\vspace{-0.5cm}

\section{Method}
\label{MET}

\subsection{MIR observational polarimetric techniques}
\label{MET_Dic}

To measure the polarization properties of an electromagnetic wave, it is necessary to separate the incident radiation in intensity modulations which are measurable by a detector. These measured components contain the polarization information of the incident radiation, which needs to be reconstructed using specific techniques. The typical instrumental components to obtain polarimetric sensitive observations is a half-wave retarder (half-wave plate, HWP) and a fixed analyzer (i.e. Wollaston prism).  The HWP rotates the incoming polarization into a set of position angles, that later is set to a constant position angle by the analyzer. In the case of the Wollaston prism, the incoming polarization, after passing through a HWP, is split in two rays with orthogonal planes of polarization. This instrumental polarization optimizes the polarization observations by obtaining instantaneous measurements of orthogonal components of polarization. These polarization components are independent of the atmospheric transmission and emission, seeing and pointing errors.

As noted in the introduction, MIR polarization observations \citep[][Lopez-Rodriguez et al. 2015 submitted]{Smith:2000aa,Packham:2007aa,Barnes:2015aa} have shown that the dominant polarization mechanism of several objects is due to dust absorption/emission of silicate-like dust grains. To obtain the absorption/emission polarization components, we followed the approach by \cite{Donn:1966aa,Dyck:1973aa}, which has been successfully applied to several objects  \citep[e.g.][]{Dennison:1977aa,Aitken:1986aa,Smith:2000aa}. Specifically, the approach by \citet{Aitken:2004aa} to obtain the normalized Stokes parameters as a function of the emissive and absorptive polarization components is followed. The absorptive polarization can be written as a function of the extinction along the long and short axis of the dust grains in the x-direction as:
\noindent
\begin{equation}
p_{\mbox{\tiny a}} = \frac{e^{-\tau_{\mbox{\tiny x}}} - e^{-\tau_{\mbox{\tiny y}}}}{e^{-\tau_{\mbox{\tiny x}}} + e^{-\tau_{\mbox{\tiny y}}}} = -\tanh \left( \frac{\Delta\tau}{2} \right) \simeq -\frac{\Delta\tau}{2}
\label{eq1}
\end{equation}
\noindent
where $\Delta\tau = (\tau_{\mbox{\tiny x}} - \tau_{\mbox{\tiny y}})$, when $\Delta\tau$ is small. The negative sign indicates that the absorptive polarization is in the direction of least extinction. Here, the polarization is orthogonal to the axis of greatest principal momentum of inertia of the dust grains. The emissive polarization can be written as:
\noindent
\begin{equation}
p_{\mbox{\tiny e}} =  \frac{\tau_{\mbox{\tiny x}} - \tau_{\mbox{\tiny y}}}{\tau_{\mbox{\tiny x}} + \tau_{\mbox{\tiny y}}} = \frac{\Delta\tau/2}{\tau} \simeq -\frac{p_{\mbox{\tiny a}}}{\tau}
\label{eq2}
\end{equation}
\noindent
where the negative sign indicates that emissive and absorptive polarization are orthogonal to each other.

The model assumes an emission source, $I_{\mbox{\tiny e}}$, that can be either unpolarized or polarized, $p_{\mbox{\tiny e}}$ at $\theta_{\mbox{\tiny e}}$, viewed through a cold dichroic sheet, $p_{\mbox{\tiny a}}$ at $\theta_{\mbox{\tiny a}}$. In this case the Stokes parameters can be written as:
\noindent
\begin{equation}
\left( \begin{matrix}
I_{\mbox{\tiny o}} \\
Q_{\mbox{\tiny o}} \\
U_{\mbox{\tiny o}} \end{matrix} \right)  = 
k \left( \begin{matrix}
1 & p_{\mbox{\tiny a}}\cos{2\theta_{\mbox{\tiny a}}} & p_{\mbox{\tiny a}}\sin{2\theta_{\mbox{\tiny a}}} \\
p_{\mbox{\tiny a}}\cos{2\theta_{\mbox{\tiny a}}} & 1-p_{\mbox{\tiny a}}\sin^{2}{2\theta_{\mbox{\tiny a}}} & p_{\mbox{\tiny a}}\cos{2\theta_{\mbox{\tiny a}}}\sin{2\theta_{\mbox{\tiny a}}} \\
p_{\mbox{\tiny a}}\sin{2\theta_{\mbox{\tiny a}}} & p_{\mbox{\tiny a}}\cos{2\theta_{\mbox{\tiny a}}}\sin{2\theta_{\mbox{\tiny a}}} & 1-p_{\mbox{\tiny a}}\cos^{2}{2\theta_{\mbox{\tiny a}}} \end{matrix} \right) 
\left( \begin{matrix}
I_{\mbox{\tiny e}} \\
Q_{\mbox{\tiny e}} \\
U_{\mbox{\tiny e}} \end{matrix} \right)
\label{eq3}
\end{equation}
\noindent
\citep[e.g.][]{S1962,Aitken:2004aa} where $I_{\mbox{\tiny o}}$, $Q_{\mbox{\tiny o}}$, and $U_{\mbox{\tiny o}}$ are the observed Stokes parameters of the object, $k$ is a factor that takes account of extinction and polarization on intensity, and $I_{\mbox{\tiny e}}$, $Q_{\mbox{\tiny e}}$, and $U_{\mbox{\tiny e}}$ are the Stokes parameters of the emissive component. 

Considering that the MIR dichroic polarization is typically smaller than 10 per cent \citep[e.g.][Lopez-Rodriguez et al. 2015, submitted]{Smith:2000aa,Barnes:2015aa}, the cross-products are negligible and
\noindent
\begin{equation}
\left(\begin{matrix}
I_{\mbox{\tiny o}} \\
Q_{\mbox{\tiny o}} \\
U_{\mbox{\tiny o}} \end{matrix} \right)  = k
\left(\begin{matrix}
I_{\mbox{\tiny e}} \\
I_{\mbox{\tiny e}}p_{\mbox{\tiny a}}\cos{2\theta_{\mbox{\tiny a}}} + Q_{\mbox{\tiny e}} \\
I_{\mbox{\tiny e}}p_{\mbox{\tiny a}}\sin{2\theta_{\mbox{\tiny a}}} + U_{\mbox{\tiny e}} \end{matrix} \right) 
\label{eq4}
\end{equation}

Using normalized Stokes parameters, $q = Q/I$ and $u = U/I$, then
\noindent
\begin{equation}
\left(\begin{matrix}
q_{\mbox{\tiny o}} \\
u_{\mbox{\tiny o}} \end{matrix} \right)  =
\left( \begin{matrix}
p_{\mbox{\tiny a}}\cos{2\theta_{\mbox{\tiny a}}} + q_{\mbox{\tiny e}} \\
p_{\mbox{\tiny a}}\sin{2\theta_{\mbox{\tiny a}}} + u_{\mbox{\tiny e}} \end{matrix} \right) =
\left( \begin{matrix}
q_{\mbox{\tiny a}} + q_{\mbox{\tiny e}} \\
u_{\mbox{\tiny a}} + u_{\mbox{\tiny e}} \end{matrix} \right)
\label{eq5}
\end{equation}
\noindent
or
\noindent
\begin{equation}
q_{\mbox{\tiny o}}(\lambda) = q_{\mbox{\tiny a}}(\lambda) + q_{\mbox{\tiny e}}(\lambda) \\
u_{\mbox{\tiny o}}(\lambda) = u_{\mbox{\tiny a}}(\lambda) + u_{\mbox{\tiny e}}(\lambda) 
\label{eq6}
\end{equation}

The observed Stokes parameters are linear functions of the absorptive and emissive Stokes parameters. If the Stokes parameters are normalized, such as $f_{\mbox{\tiny a}}(\lambda) = p_{\mbox{\tiny a}}(\lambda)/p_{\mbox{\tiny a}}(\lambda_{\mbox{\tiny max}})$, and $f_{\mbox{\tiny e}}(\lambda) = p_{\mbox{\tiny e}}(\lambda)/p_{\mbox{\tiny e}}(\lambda_{\mbox{\tiny max}})$, then $q_{\mbox{\tiny a}} \propto f_{\mbox{\tiny a}}$ and $q_{\mbox{\tiny e}} \propto f_{\mbox{\tiny e}}$, and Eq. (\ref{eq6}) can be written as:
\noindent
\begin{equation}
q_{\mbox{\tiny o}}(\lambda) =  Af_{\mbox{\tiny a}}(\lambda) + Bf_{\mbox{\tiny e}}(\lambda)\\
u_{\mbox{\tiny o}}(\lambda) = Cf_{\mbox{\tiny a}}(\lambda) + Df_{\mbox{\tiny e}}(\lambda) 
\label{eq7}
\end{equation}
\noindent
where $A,B,C,D$ are constant factors of the normalized emissive/absorptive components $f_{\mbox{\tiny e}}(\lambda)$ and $f_{\mbox{\tiny a}}(\lambda)$, that need to be estimated through a fitting procedure.

Following Eq. (\ref{eq2}), $p_{\mbox{\tiny e}}(\lambda) = p_{\mbox{\tiny a}}(\lambda)/\tau_{\lambda}$, the  emissive component, $f_{\mbox{\tiny e}}(\lambda)$, is a function of the extinction curve, $\tau_{\lambda}$. This condition can be used as long as the difference in orthogonal optical depths of the dust grains is less than unity for $p_{\mbox{\tiny a}}$, which makes $p_{\mbox{\tiny a}}$ independent of the optical depth. This situation will hold if $\tau_{\mbox{\tiny MIR}} <$ a few tens. The appropriate extinction curve will depend on the object, i.e. active galactic nuclei, star forming regions, etc. The profile of the absorptive component depends on the dust composition of the astrophysical environment. Although the absorptive component of the Becklin-Neugebauer (BN) object in Orion \citep{Aitken:1989aa} has been successfully applied  \citep[e.g.][]{Dennison:1977aa,Aitken:1986aa,Smith:2000aa} to represent silicate-like polarization profiles, other absorptive components should be considered for non-silicate astrophysical environments (e.g. Lopez-Rodriguez et al. 2015, submitted). It is out of the scope of this paper to define the absorptive and emissive components of polarization for non-silicate dust grains. However, the model presented in this paper allows the user to introduce any profile for the absorptive/emissive components. Thus, the developed Bayesian model (Section \ref{MET_Bayes}) offers the possibility to distinguish between dust composition and extinction curves of any astrophysical objects by the customization of the absorptive/emissive components.

\subsection{Bayesian approach}
\label{MET_Bayes}

The model presented by \citet{Aitken:2004aa} is optimized to minimize $\chi^{2}$ for the normalized Stokes parameters $q$ and $u$, independently. This method provides a unique best-fit of the observations regardless of whether or not a unique solution exists, nor whether the Stokes parameters are dependent on each other. In this paper we focus on simultaneously inferring the normalized Stokes parameters to find the full family of possible solutions rather than focusing on the best-fit estimate. 

Let $\theta = \{\theta_{\tiny 1}, \dots, \theta_{\tiny n} \}$ represent the set of unknown model parameters and let $\bld{d} = \{d_{\tiny 1}, \dots, d_{\tiny n} \}$ represent a set of observed data. The probability for an observation to occur under a given set of parameters is given by the likelihood model $P(\bld{d}|\theta)$. As our aim is to obtain the family of solutions that best explain the observations, the likelihood model can be written as  $P(\bmath{\theta}|\bld{d})$, called the posterior distribution. Thus, we can construct the Bayes theorem as $P(\bmath{\theta}|\bld{d}) = (P(\bld{d}|\theta) \times P(\theta))/P(\bld{d})$, where $P(\theta)$ is the prior distribution, and $P(\bld{d})$ is a normalization constant that does not affect the shape of the posterior distribution and thus can be ignored, allowing us to rewrite the posterior distribution as $P(\bmath{\theta}|\bld{d}) \propto P(\bld{d}|\theta) \times P(\theta)$, or simply $P( \bmath{\theta}|\bld{d}) \propto P(\bmath{\theta}, \bld{d})$. Thus, the probabilistic distribution $P(\bmath{\theta}, \bld{d})$ needs to be set through the identification of the region of the parameter space that can explain the observations.


\begin{figure}
\includegraphics[angle=0,trim=0cm 0.8cm 0cm 0cm,scale=0.26]{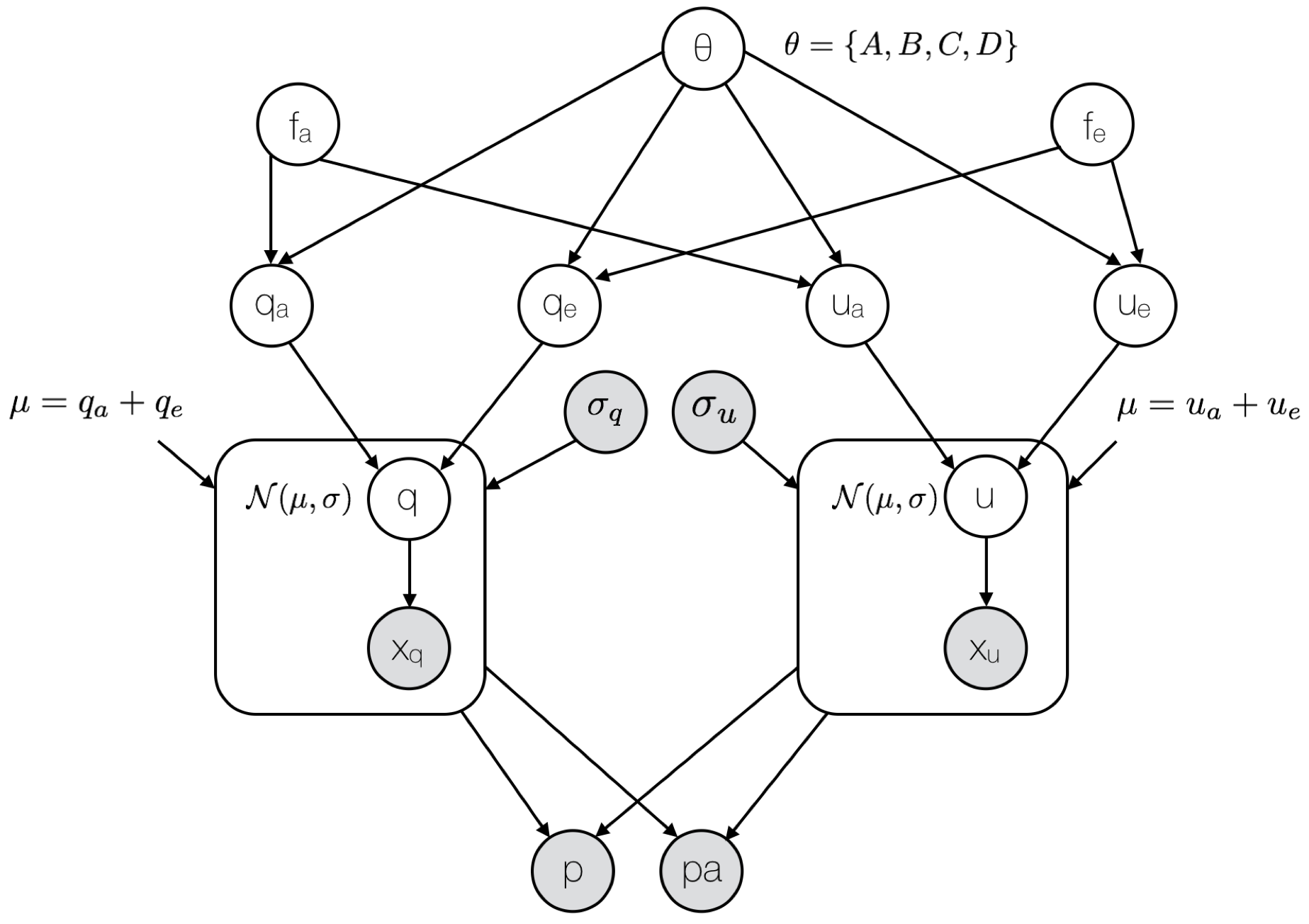}
\caption{Bayesian network graph. All modeled parameters (white circles) depend on the uniform distributions $\theta = \{A,B,C,D\}$, and the user defined absorptive, $f_{\mbox{\tiny a}}$, and emissive, $f_{\mbox{\tiny e}}$, components. The observables (grey circles) correspond to the uncertainties, $\sigma_{\mbox{\tiny q}},~\sigma_{\mbox{\tiny u}}$, of the observed normalized Stokes parameters $x_{\mbox{\tiny q}}$, $x_{\mbox{\tiny u}}$, and the degree, $p$, and angle, $pa$, of polarization. The modeled Stokes parameters $q$ and $u$ are linear combinations of the absorptive and emissive comments of each Stokes parameter, i.e. $\mu = q_{\mbox{\tiny a}} + q_{\mbox{\tiny e}}$. $\mu$ is assumed to be a Normal distribution for each polarimetric data per wavelength. Once the model obtains the inference of $q$, and $u$, the degree and angle of polarization are estimated.}
\label{fig1}
\end{figure}


Let us define the set of unknown parameters $\theta$ and set the observables $\bld{d}$. Based on Eq. (\ref{eq7}), the set of free parameters are $A$, $B$, $C$, and $D$, thus $\theta = \{ A, B, C, D\}$, and the user defined absorptive, $f_{\mbox{\tiny a}}$, and emissive, $f_{\mbox{\tiny e}}$, components. For the set of observations, we adopt the polarimetric observations (imaging- and/or spectro-polarimetric) as characterized by the normalized Stokes parameters and their uncertainties $\bld{d} = \{q, u, \sigma_{\tiny q}, \sigma_{\tiny u} \}$. Fig. \ref{fig1} shows the Bayesian network as a hierarchical regression where the free parameters are $\theta = \{ A, B, C, D\}$, and the user defined absorptive, $f_{\mbox{\tiny a}}$, and emissive, $f_{\mbox{\tiny e}}$, components. Each node corresponds to a variable from $\theta$ or $\bld{d}$, and the user defined absorptive, $f_{\mbox{\tiny a}}$, and emissive, $f_{\mbox{\tiny e}}$, components, represented as white circles in Fig. \ref{fig1}. The observational data is represented as grey circles, i.e. observed Stokes parameters shown as $x_{\mbox{\mbox{\tiny q}}}$ and $x_{\mbox{\mbox{\tiny u}}}$, and the degree, $p$, and  angle, $pa$, of polarization. Based on Eq. (\ref{eq6}), the modeled Stokes parameters $q$ and $u$, are linear combinations of the emissive/absorptive components, represented as $\mu$ in Fig. \ref{fig1}. 

 We took the prior distributions $P(\theta)$ as uniform functions with values five times larger and lower than the maximum and minimum value of the observed Stokes parameters $q$ and $u$, i.e. $P(\theta) \sim \mathcal{U}(\min,\max) = \mathcal{U}(5 \times \min(\theta), 5 \times \max(\theta))$. As the observed Stokes parameters, $q$ and $u$, are linear combinations of the emissive/absorptive components and each observable data point has an associated uncertainty, $\sigma$, the prior is assumed to be a normal distribution, i.e. Gaussian function, given by $P(\bmath{\theta}, \bld{d}) = \mathcal{N}(\mu,\sigma)$. Once the family of solutions are determined for the normalized Stokes parameters $q$ and $u$, the posterior distributions are used to estimate the degree, $p = \sqrt{q^{2} + u^{2}}$, and angle, $pa = 0.5\arctan{(u/q)}$, of polarization.

Several libraries allowing Bayesian models using Markov Chain Monte Carlo (MCMC) algorithms are available. In this paper, we use the {\small PyMC3}\footnote{{\small PyMC3} is available at \url{https://github.com/pymc-devs/pymc3}} framework for Python, which has been successfully applied to a variety of astrophysical problems \citep[e.g.][]{Genet:2010aa,Barentsen:2013aa,Wilkins:2013aa,Waldmann:2014aa}. {\small PyMC3} is a Python module for Bayesian statistical modeling and model inference using MCMC algorithms. This Python module uses the No-U-Turn Sampler \citep[NUTS;][]{Hoffman:2011aa}, a Hamiltonian MCMC that avoids the random walk behavior and sensitivity of other MCMC algorithms by taking a series of steps through a first-order gradient information. The implementation of this algorithm allows a high-dimensional target distribution to converge more quickly than other methods, such as Metropolitan or Gibbs sampling. This paper contains a precise and repeatable specification of the parameters estimation procedure. The source code and accompanying files are available at the GitHub repository of the author\footnote{The full Python code is available at \url{https://github.com/enloro/MIR\_Pol\_Bayes}}.

\subsection{Example: W51 IRS2}
\label{MET_Example}

\begin{figure}
\includegraphics[angle=0,trim=0cm 1cm 0cm 0cm,scale=0.50]{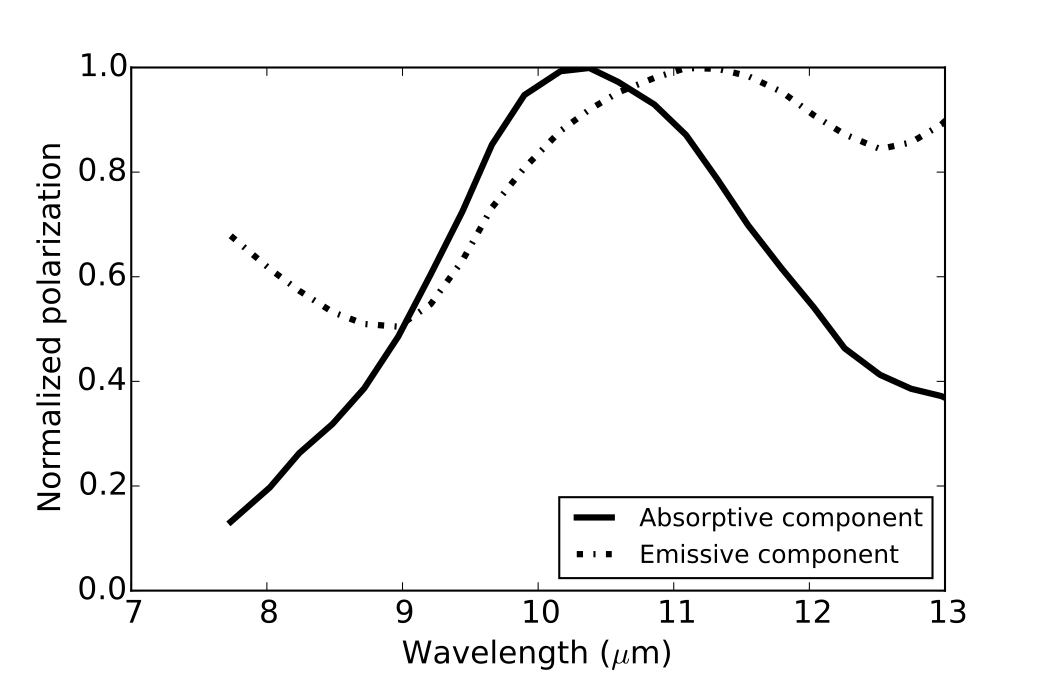}
\caption{Normalized absorptive, $f_{\mbox{\tiny a}}$($\lambda$), and emissive, $f_{\mbox{\tiny e}}$($\lambda$), profiles. The absorptive profile is the normalized BN polarization spectrum, and the emissive profile is $f_{\mbox{\tiny a}}$($\lambda$)$\sim f_{\mbox{\tiny a}}$($\lambda$)/$\tau_{\lambda}$.}
\label{fig2}
\end{figure}

\begin{figure*}
\includegraphics[angle=0,trim=0.5cm 0.8cm 0cm 0cm,scale=0.4]{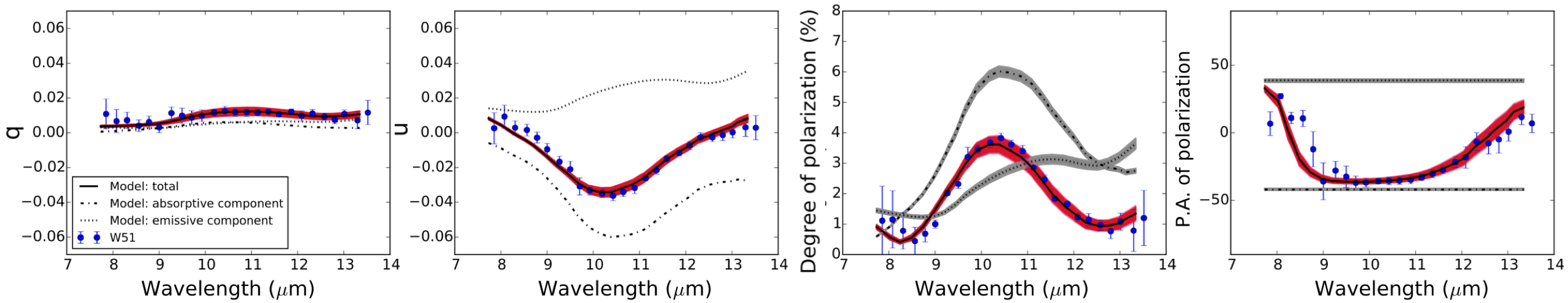}
\caption{Spectropolarimetric observations (blue dots) of W51 IRS2 \citep{Smith:2000aa} and the inference model from the Bayesian approach. From left to right: the normalized $q$ and $u$ Stokes parameters, degree and P.A. of polarization. In each plot, the absorptive (dashed-dotted black line) and emissive (dotted black line) components, and the total (solid black line) model are shown. The 95\% (2$\sigma$) HDP of the total (red shadowed regions), absorptive and emissive (grey shadowed regions) components are shown.}
\label{fig3}
\end{figure*}


To verify the reliability of the Bayesian approach presented in Section \ref{MET_Bayes}, the spectropolarimetric observations of W51 IRS2 taken from \citet{Smith:2000aa} were used. This set of observations was used by \citet{Aitken:2004aa} on their $\chi^{2}$ best-fit approach allowing us a direct comparison of both methods. W51 IRS2 was observed with the University College of London (UCL) spectropolarimeter mode in the 8$-$13 \um~wavelength range with a 4.3 arcsec beam size on the 3.8-m United Kingdom Infrared Telescope (UKIRT). To apply the developed Bayesian approach in Section \ref{MET_Bayes}, we need to define the absorptive and emissive components, as well as the extinction curve. For the absorptive component, $f_{\mbox{\tiny a}}$, we took the BN object in Orion \citep{Aitken:1989aa} because it is the best-defined absorptive component with the highest signal-to-noise MIR spectro-polarimetric observations. Following Eq. (\ref{eq2}),  the normalized emissive component is a function of the extinction curve, $\tau_{\lambda}$. We have taken the extinction curve due to silicates, derived from the Trapezium region of Orion \citep{Gillett:1975aa}. Fig. \ref{fig2} shows the absorptive and emissive components used as inputs in the model. Note that the emissive component estimated through this procedure is similar to the observed emissive profile of SgrA IRS1 \citep[][fig. 2]{Smith:2000aa}. 

Using the developed Bayesian approach, the sampling was carried out using four free parameters, $A$, $B$, $C$, and $D$ (Section \ref{MET_Dic}) of 25000 steps (with a burn-in length of 5000 steps), and the user defined absorptive/emissive components. These independent chains are found to converge to the same parameter-space regions within a few hundred iterations, meaning that a global maximum was reached in all cases with a fast convergence. The most probable inference is shown in Fig. \ref{fig3} as a black solid line, where the 95\% (2$\sigma$) high distribution probability (HDP) of the family of solutions is shown as red (total model) and grey (absorptive and emissive components) shadowed regions. Table \ref{table1} shows the measured degree and angle of polarization of the absorptive and emissive components at the peak of the absorptive component, 10.7 \um, from our Bayesian approach and those from \citet{Smith:2000aa}. Although the degree of polarization are very similar, slightly differences in the polarization angle are found. The Bayesian approach is able to estimate the polarization angle with lower uncertainty than previous $\chi^{2}$ methods. This result has direct implications in the estimation of the magnetic field geometry and strength for tomographic studies \citep[i.e.][]{Aitken:1998aa,Barnes:2015aa}. For instance, if the magnetic field strength is estimated through the Chandrasekhar-Fermi method \citep{Chandrasekhar:1953aa}, the magnetic field strength, $B \propto 1/\alpha$, depends on the dispersion of the polarization angle, $\alpha$. Based on Table \ref{table1}, the magnetic field strength from $\chi^{2}$ methods can be underestimated up to a factor three in comparison with the developed Bayesian approach.


\begin{table}
\caption{Comparison with \citet{Smith:2000aa}}
\label{table1}
\begin{tabular}{ccc}
\hline
Component									&	\citet{Smith:2000aa}		&	This work	\\			
\hline	
$p_{\mbox{\tiny a}}$							&	6.0$\pm$0.3\%			&	6.0$\pm$0.2\%		\\
$pa_{\mbox{\tiny a}}^{\mbox{\tiny (a)}}$	&	136$\pm$1\degr		&	138$\pm$1\degr		\\
$p_{\mbox{\tiny e}}$							&	2.8$\pm$0.3\%			&	2.6$\pm$0.1\%		\\
$pa_{\mbox{\tiny e}}$							&	36$\pm$3\degr			&	39$\pm$1\degr		\\
\hline
\end{tabular}
\\
$^{a}$For direct comparison with \citet{Smith:2000aa}, the polarization angle is shown as $pa_{\mbox{\tiny a}}=-42\degr+180\degr = 138\degr$.
\end{table}

The change in polarization angle with wavelength suggests the contribution of several polarization components. Based on the developed Bayesian approach, if only one mechanism of polarization is present, the non-dominant component will show a flat posterior distribution. This flat profile provides the user with immediate information if both or only one polarization components is present in the object. This represents an advantage from $\chi^{2}$ methods, where it needs to be applied several times to estimate the minimum $\chi^{2}$ for emissive and absorptive components simultaneously and for each component individually, and then compare them. In the case of W51 IRS2, both emissive and absorptive polarization components are present in the final polarization. This result is in agreement with the observed dependence of the polarization angle with wavelength.

Spectropolarimetric observations are not always accessible for all objects, however imaging-polarimetry offers a solution to this problem. The selection of wavelength and number of filters is crucial to disentangle the polarization mechanisms of the object. The developed Bayesian model was run (Section \ref{A_MIR_Imapol}) using two or more simulated imaging-polarimetric observations with a bandwidth of 1 \um~in the spectral range of 8$-$13 \um. A bandwidth of 1 \um~covering the 8$-$13 \um~wavelength range is the typical filter set. The dispersion of inferred models is minimized when two filters (Fig. \ref{fig4}d,f), on the edges and central wavelength range, and three filters (Fig. \ref{fig4}g,h) covering the whole spectral range are selected. Uncertainties of 0.5 per cent and 3\degr~in the degree and angle of polarization, respectively, are found for these configurations. These uncertainties represent an improvement on the fitting technique when compared with previous $\chi^{2}$ methods, with typical uncertainties of 1 per cent and 4\degr~on the degree and angle of polarization, respectively. Thus, imaging-polarimetric observations using two or three filters located in the central 9.5$-$10.5 \um, and the edges, 8$-$9 \um~and/or 11$-$13 \um, of the wavelength range are recommended to optimally disentangle the polarization mechanisms in the 8$-$13 \um~wavelength range.

\vspace{-0.5cm}

\section{Conclusions}
\label{CON}
 
A fast and general Bayesian inference framework was presented and applied to infer the physical properties of the dichroic polarization using MIR polarimetric observations. From a statistical framework, the Bayesian approach presented here allows the user to simultaneously infer the normalized Stokes parameters to find the full family of solutions that best describe the observations. This approach allows the user to obtain 1) more precise estimations of the magnetic field strength and geometry for tomographic studies, and 2) immediate information about the dominant polarization components of the object. The Bayesian approach was run using simulated imaging-polarimetric observations to obtain the optimal filter configuration to disentangle the polarization mechanism. Imaging-polarimetric observations using two or three filters located in the central 9.5$-$10.5 \um, and the edges, 8$-$9 \um~and/or 11$-$13 \um, of the wavelength range are recommended to optimally disentangle the polarization mechanisms in the 8$-$13 \um~wavelength range. This method offers a tool for high-spatial imaging- and spectra-polarimetric observations for the current polarimeters in 10-m class and potential polarimetric capabilities in the next generation of 30-m class telescopes.

\section*{Acknowledgments}

E.L.R. would like to thank the anonymous referee for their useful comments, which improved the paper significantly. E.L.R acknowledges support from the University of Texas at San Antonio and University of Texas at Austin.

\bibliographystyle{mnras}
\bibliography{Bayes_paper_min}


\appendix 

\section{MIR imaging-polarimetry}
\label{A_MIR_Imapol}
\begin{figure*}
\includegraphics[angle=0,trim=-0.5cm 0cm 0cm 0cm,scale=0.8]{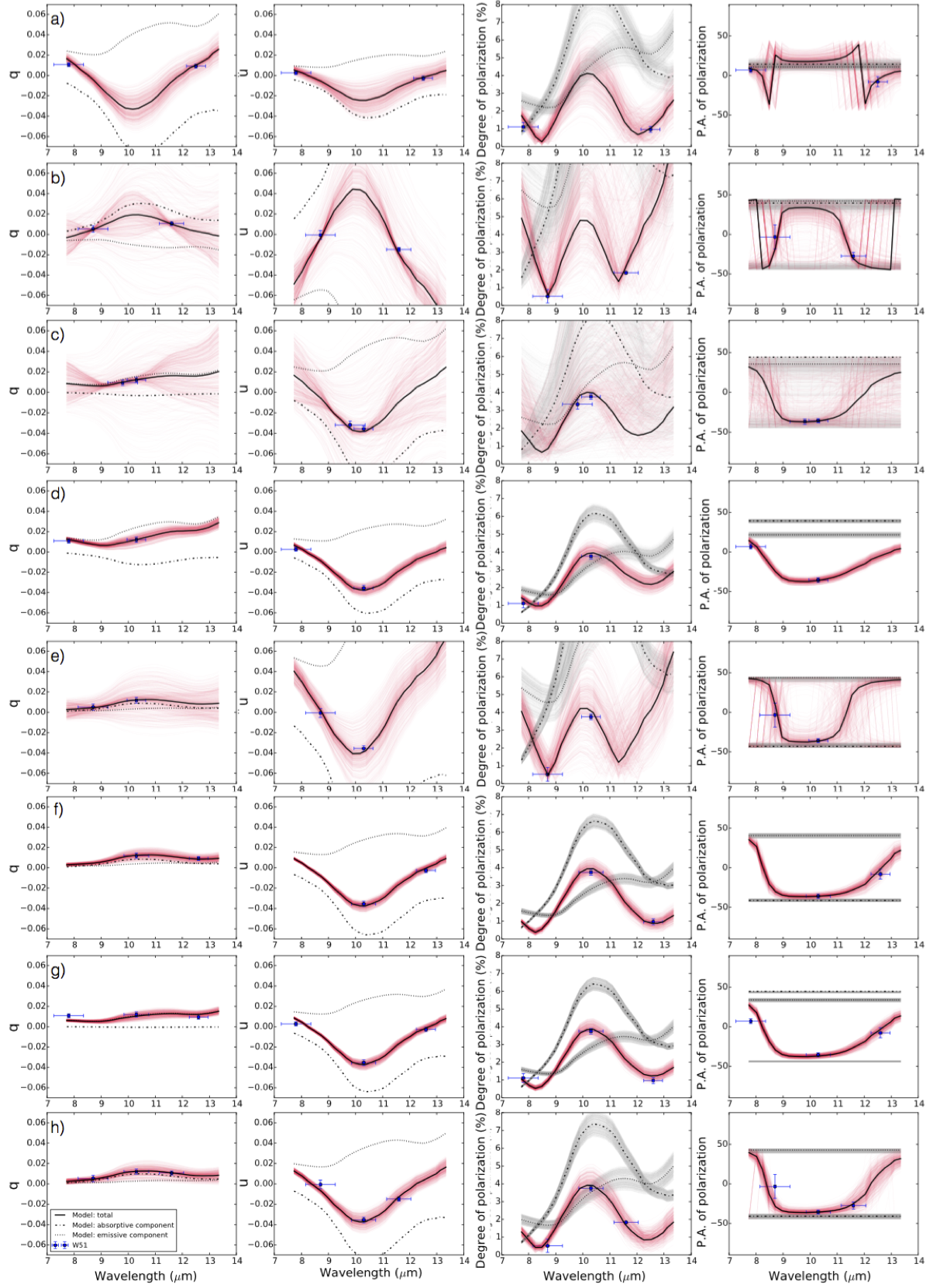}
\caption{Imaging-polarimetric simulated observations (blue dots) of W51 IRS2 \citep{Smith:2000aa} and the inference model from the Bayesian approach. Same plot configuration as Fig. \ref{fig3}. Note that the model dispersion is minimized in the cases where two filters (d and f), one on the edge and one on the central wavelength range, and three filters (g and h) are selected.}
\label{fig4}
\end{figure*}

Imaging-polarimetric simulated observations of W51 IRS1 \citep{Smith:2000aa} were used. As described in Section \ref{MET_Example}, the developed Bayesian approach was run using two or more simulated imaging-polarimetric observation with a bandwidth of 1 \um~in the spectral range of 8$-$13 \um. Figure \ref{fig4} shows the normalized Stokes parameters, degree and angle of polarization for the combination of two and three filters. Specifically, a) 7.8 \um~and 12.5 \um, b) 8.7 \um~and 11.6 \um, c) 9.7 \um~and 10.3 \um, d) 7.8 \um~and 10.3 \um, e) 8.7 \um~and 10.3 \um, f) 10.3 \um~and 12.5 \um, g) 7.5 \um, 10.3 \um~and 12.5 \um, and h) 8.7 \um, 10.3 \um~and 11.6 \um. In all cases, the Bayesian approach was carried out using 5100 steps (with a burn-in of 100 steps), with the same prior distributions $P(\theta)$. The maximum and minimum of the Stokes parameters were $u =[-0.2,0.03]$ and $q=[-0.1,0.2]$.

\bsp	
\label{lastpage}
\end{document}